\documentclass[conference]{IEEEtran}
\IEEEoverridecommandlockouts
\usepackage{booktabs}
\usepackage{multirow}
\usepackage{flushend}
\usepackage{cite}
\usepackage{amsmath,amssymb,amsfonts}
\usepackage{algorithmic}
\usepackage{graphicx}
\usepackage{textcomp}
\usepackage{xcolor}
\def\BibTeX{{\rm B\kern-.05em{\sc i\kern-.025em b}\kern-.08em
    T\kern-.1667em\lower.7ex\hbox{E}\kern-.125emX}}

\begin{document}
% % \documentclass[conference]{IEEEtran}
% \IEEEoverridecommandlockouts
% % The preceding line is only needed to identify funding in the first footnote. If that is unneeded, please comment it out.
% \usepackage{booktabs}
% \usepackage{multirow}
% \usepackage{flushend}
% \usepackage{cite}
% \usepackage{amsmath,amssymb,amsfonts}
% \usepackage{algorithmic}
% \usepackage{graphicx}
% \usepackage{textcomp}
% \usepackage{xcolor}
% \def\BibTeX{{\rm B\kern-.05em{\sc i\kern-.025em b}\kern-.08em
%     T\kern-.1667em\lower.7ex\hbox{E}\kern-.125emX}}
% \begin{document}

\makeatletter

\def\ps@IEEEtitlepagestyle{%
  \def\@oddfoot{\mycopyrightnotice}%
  \def\@evenfoot{}%
}
\def\mycopyrightnotice{%
  {\footnotesize 978-1-6654-6819-0/22/\$31.00 \copyright2022 IEEE\hfill}
  \gdef\mycopyrightnotice{}
}

\title{Finding the Most Transferable Tasks for Brain Image Segmentation\thanks{This research is funded by Natural Science Foundation of China 62001266. Supplementary materials are available in this arXiv version.}
}

\author{\IEEEauthorblockN{Yicong Li, Yang Tan, Jingyun Yang, Yang Li, Xiao-Ping Zhang}
\IEEEauthorblockA{\textit{Tsinghua-Berkeley Shenzhen Institute}\\
\textit{Tsinghua University}\\
Shenzhen, China\\
yangli@sz.tsinghua.edu.cn}}

\maketitle

\begin{abstract}
Although many studies have successfully applied transfer learning to medical image segmentation, very few of them have investigated the selection strategy when multiple source tasks are available for transfer. In this paper, we propose a prior knowledge guided and transferability based framework to select the best source tasks among a collection of brain image segmentation tasks, to improve the transfer learning performance on the given target task. The framework consists of modality analysis, RoI (region of interest) analysis, and transferability estimation, such that the source task selection can be refined step by step. Specifically, we adapt the state-of-the-art analytical transferability estimation metrics to medical image segmentation tasks and further show that their performance can be significantly boosted by filtering candidate source tasks based on modality and RoI characteristics. Our experiments on brain matter, brain tumor, and white matter hyperintensities segmentation datasets reveal that transferring from different tasks under the same modality is often more successful than transferring from the same task under different modalities. Furthermore, within the same modality, transferring from the source task that has stronger RoI shape similarity with the target task can significantly improve the final transfer performance. And such similarity can be captured using the Structural Similarity index in the label space.
\end{abstract}


\vspace{-0.5cm}
\section{Introduction}
Supervised deep learning algorithm requires an abundant amount of annotated data to obtain a well-trained model that can make accurate predictions. Such a requirement severely limits the application of deep learning in the medical domain since acquiring and labeling medical data can be very expensive and time-consuming. A commonly used solution is transfer learning \cite{pan2009survey}: pre-training a model on a source task where sufficient annotated data exists and fine-tuning the model on the desired target task where only a small amount of annotated data is available. A key question to the success of transfer learning between source and target tasks is what source task shall we transfer from given a target task (i.e., what to transfer). That is, if we transfer knowledge from a less related source task, it may inversely hurt the performance on the target task, a phenomenon known as negative transfer \cite{wang2019characterizing}. Although many studies have already successfully applied transfer learning to medical image analysis problems \cite{huang2021cross,kermany2018identifying,chen2019med3d}, very few of them have investigated how to select the best source tasks. Hence in this work, we aim to systematically tackle the {\em source selection problem}, focusing particularly on the most dominant image segmentation tasks. 

An effective source selection strategy requires a good understanding of what factors affect the transfer performance for medical segmentation tasks. Several experimental studies have investigated the transfer learning performance in medical image analysis. In \cite{wen2020effective}, advantages and disadvantages of several transfer learning strategies for medical image segmentation tasks are explored, such as which component of a CNN model is better to transfer. Weatheritt et al. \cite{weatheritt2020transfer} analyze the impacts of different choices of data pre-processing methods, tasks, and amount of data used on the final transfer performance. Unfortunately, these studies haven't fully investigated how detailed characteristics of source tasks can influence transfer learning performance. Particularly, for medical image segmentation, image characteristics of source tasks such as the modality and the geometric shape of the segmentation region of interest (RoI) vary greatly due to the fundamental differences in scanner protocols, imaging procedures, lesion positions, etc. Such variations will definitely affect the final transfer learning performance on the target task.
% Thus, we firstly provide a comprehensive study of the relationship between image characteristics (modality and RoI shape) and transfer learning performance to explore meaningful insights on how to select appropriate sources, using brain matter segmentation, brain tumor segmentation, and white matter hyperintensities segmentation datasets.

% Taking a step further, because of the aforementioned heterogeneity that lies in different medical image segmentation datasets, it is very hard to predict which pre-trained source model will perform better after fine-tuning on target task when multiple source tasks are available. One naive option is actually conducting pre-training and fine-tuning experiments for all source tasks, which is quite time-consuming and subject to constraints of computing resources.
To tackle the source selection problem systematically, we need a quantitative way to rank the transfer performance of source tasks. Recent studies have proposed different methods to estimate the {\em knowledge transferability} between source and target tasks for natural images. Transferability reveals how easy it is to transfer knowledge learned from a source task to a target task. Bao et al. \cite{bao2019information} takes an information theoretic approach and develops a computable metric called H-score to estimate the knowledge transferability between datasets for image classification problems. Under the assumption that the inputs of source and target tasks share the same domain, Tran et al. \cite{tran2019transferability} uses NCE, negative conditional entropy, to evaluate transferability. In \cite{nguyen2020leep}, a new metric called LEEP is proposed. It makes predictions based on the expected empirical conditional distribution between source and target labels. More recently, Tan et al. \cite{tan2021otce} proposes an optimal transport based conditional entropy (OTCE) metric to analytically predict transfer performance for supervised classification problems, which has been further extended to evaluate segmentation problems as well \cite{tan2021transferability}. However, these works focus on finding a general-purpose transferability metric purely based on feature effectiveness without consideration of prior knowledge (e.g., image characteristics) about tasks, such as modality difference and RoI shape similarity between source and target tasks in the medical image domain. 
% And also there hasn't been a study that comprehensively evaluates the performance of these metrics on medical image segmentation tasks even though huge differences exist between medical images and natural images.

As such, in this work, we propose a novel source selection framework that leverages the prior knowledge of medical image segmentation tasks for reducing the computation time and improving the selection accuracy of existing transferability metrics, so as to create a systematic way to select the best source tasks for a given target task. We choose brain image segmentation datasets for our experiments since it is one of the most challenging and time-consuming clinical procedures for diagnosing brain disorders, whose demand has been increasing in recent years. In summary, our main contributions are:
\begin{itemize}
    \item \textbf{A prior knowledge guided and transferability based source selection framework.} The framework incorporates prior knowledge of medical image segmentation tasks with transferability estimation metrics to select the best source tasks for transfer learning given a target task.
    % Experiments show that with the enhancement of prior knowledge, the performance of current state-of-the-art transferability estimation metrics can be significantly boosted.
    \item \textbf{An analysis of the relationship between image characteristics and transfer learning performance.} We extensively conduct transfer learning experiments under the cross-modality or cross-task setting and conclude that transferring from a different task under the same modality is often more beneficial than transferring from the same task under a different modality. We also quantitatively show that within the same modality, transferring from the source task that has stronger RoI shape similarity with the target task outperforms transferring from those less similar ones.
% These findings are examined on three publicly available benchmark datasets including brain matter segmentation, brain tumor segmentation, and white matter hyperintensities segmentation.
\end{itemize}
% (1) A prior knowledge guided and transferability based source selection framework; (2) An analysis of the relationship between image characteristics and transfer learning performance.

\section{Methods}
\subsection{Problem Description}
In a typical source selection problem, we are given $K$ pre-trained source models $\{\theta_s\}^K_{k=1}$ corresponding to $K$ source task $\{T_s\}^K_{k=1}$ and a target task $T_t$. The problem goal is to find out which source task $T_s$  we should transfer from in order to achieve the best transfer performance on the target task $T_t$.

The most common source selection approach is to fine-tune each source model on the target task to obtain a transfer accuracy on the target test set. This transfer accuracy is called the \textit{ground truth transferability}, which can be represented by a certain segmentation accuracy evaluation metric, such as Dice score. Then the task that corresponds to the source model that achieves the best transfer accuracy will be selected as the most appropriate source. However, this naive method is very computationally expensive and may become very inefficient when $K$ is large. Therefore, analytical methods are developed to estimate the transferability without fine-tuning each source model on the target task to save time and resources. The ranking of transferability scores between source and target tasks produced by such analytical methods should well correlate with the ranking given by the ground truth transferability scores so that top-performing sources can be selected accurately but in a less time-consuming way.

\subsection{Prior Knowledge Guided and Transferability Based Source Selection Framework}
\begin{figure}[t]
  \centering
  \includegraphics[width=1\linewidth]{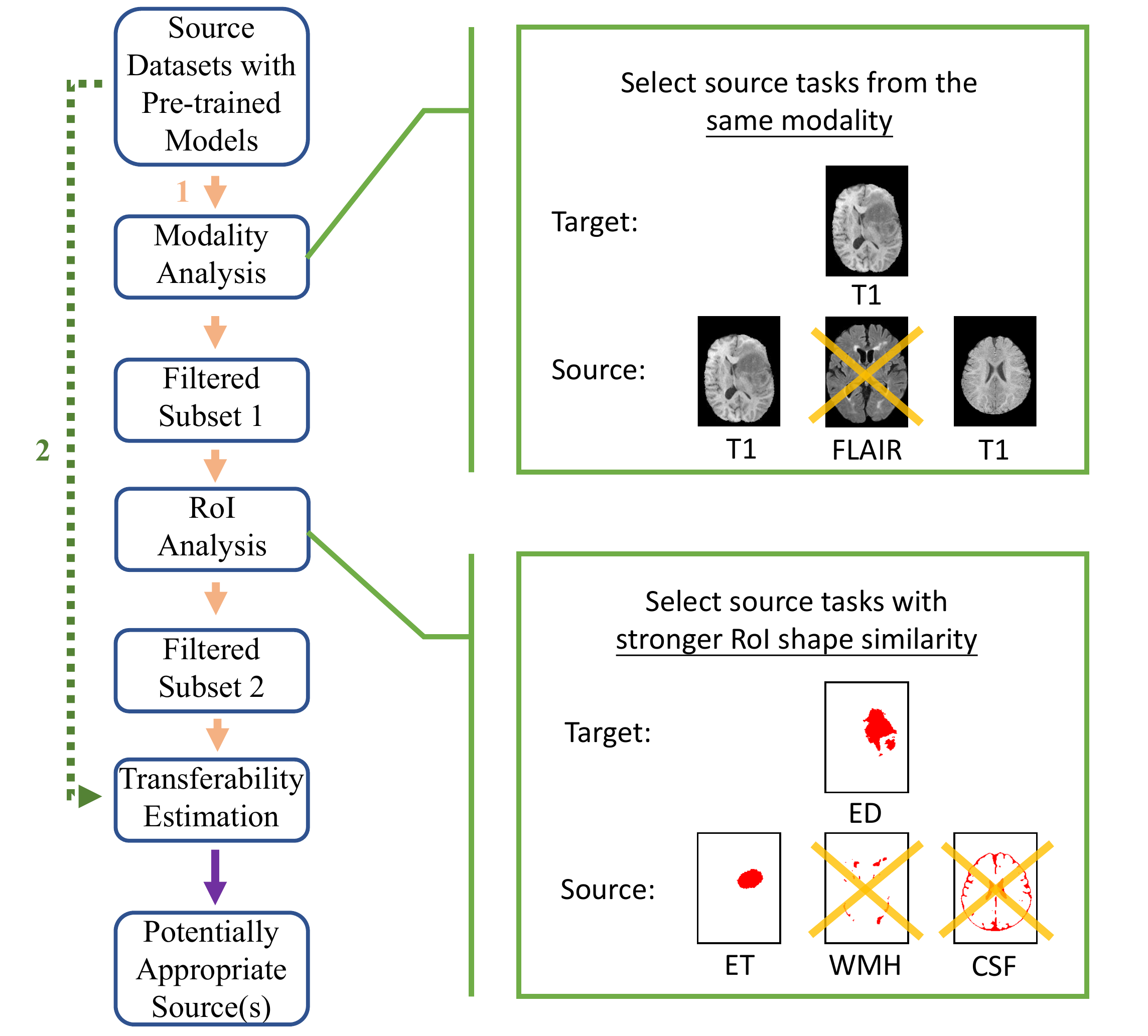} %pipeline_real_image.pdf}
  \caption{Prior knowledge guided and transferability based source selection framework. Path 1: do modality analysis and RoI analysis, then use transferability estimation metrics. Path 2: apply transferability estimation metrics directly on the source tasks. T1: t1-weighted modality, FLAIR: fluid-attenuated inversion recovery modality, ED: edematous tissue, ET: enhancing tumor, WMH: white matter hyperintensities, CSF: cerebrospinal fluid.}
  \label{fig:sources_selection_framwork}
  \vspace{-0.5cm}
\end{figure}

Unfortunately, analytical transferability estimation metrics like H-score and OTCE are designed for natural images, thus their performance in the medical image domain is not guaranteed. Inspired by the differences in modalities and RoI shapes of medical image segmentation tasks, we propose a source selection framework that incorporates the analysis of image characteristics with current state-of-the-art transferability estimation metrics, as shown in Fig. \ref{fig:sources_selection_framwork}.

Given a pool of source tasks and a target task, we propose to analyze the image characteristics of tasks before computing transferability estimation metrics (Path 1 in Fig. \ref{fig:sources_selection_framwork}). Specifically, our framework consists of three steps:
\subsubsection{\textbf{Modality Analysis}}
\label{sec:modality}
Select source tasks that are under the same modality as the target task to generate Subset 1 (we define a subset as a smaller group of tasks selected from a bigger group of tasks).

\subsubsection{\textbf{RoI Analysis}}
\label{sec:roi}
Within Subset 1 (in which source tasks are under the same modality as the target task), select source tasks whose RoI shapes are more similar to that of the target task by calculating the shape similarity to generate Subset 2. Specifically, we propose to use structural similarity index measure (SSIM) \cite{wang2004image} to quantify the RoI shape similarity. In our transfer learning context, given a source task $T_s$ with its label set $Y_s$ and a target task $T_t$ with its label set $Y_t$, we compute their RoI shape similarity as:
\begin{equation}
\begin{small}
  \textit{RoI-Sim}(T_s, T_t) = \textit{SSIM}(Y_s, Y_t).
\end{small}
\end{equation}
More details on SSIM can be found in the supplementary materials.

\subsubsection{\textbf{Transferabiltiy Estimation}}
\label{sec:transferability}
Within Subset 2, we apply a certain analytical transferability estimation metric to select potentially appropriate source tasks. In this work, we choose H-score \cite{bao2019information} or OTCE \cite{tan2021transferability} as the metric.
% Suppose we have a source dataset $D_s = {\{(x_s^i, y_s^i)\}}_{i=1}^m$ with $m$ samples and a target dataset $D_t = {\{(x_t^i, y_t^i)\}}_{i=1}^n$ with $n$ samples, where $x$ and $y$ denote the image and the label of a sample, we can pre-train a source model $\theta_s$ on the source dataset $D_s$. Then we test the source model $\theta_s$ with $D_s$ and $D_t$ to get source and target features, denoted as $F_s = \{f_s^i, y_s^i\}_{i=1}^m$ and $F_t = \{f_t^i, y_t^i\}_{i=1}^n$ respectively, where $f$ is the extracted feature of an input image $x$.

% To compute H-score, we need to directly test the source model $h_S$ using target data $D_t$ to get target feartures $F_t$. Then, final H-score for a $(D_s, D_t)$ pair is computed as:
% \vspace{-0.3cm}
% \begin{equation}
% \vspace{-0.3cm}
% \begin{aligned}
% \textit{H-score}(D_s, D_t) = \frac{1}{n}\sum_{i=1}^{n}\frac{1}{N}\sum_{j=1}^{N}M_\textit{H-score}(f_t^i, y_t^{i, j})
% \end{aligned}
% \label{eq:hgr}
% \end{equation}
% where $N$ is number of pixels in the label image and $y^{i, j}$ is the label for the $jth$ pixel in the $ith$ image. Higher H-score indicates better transfer performance after fine-tuning. More details of the core function $M_\textit{H-score}$ can be found in \cite{bao2019information}.

%To compute H-score,  
In the task transfer learning setting, let $(X_s, Y_s)$ and $(X_t, Y_t)$ represent the input and output random variables of the source and target tasks, respectively.  %H-score assumes that $P(X_s)=P(Y_t)$ but $P(Y_s|X_s)=P(Y_t|X_t)$ .  
Given a source model $\theta_s$ pre-trained on a source task, we denote the feature of source and target data $X_s$, $X_t$ extracted by the source model $\theta_s$ as $\theta_s(X_s)$ and $\theta_s(X_t)$. Assuming the same input domain $P(\theta_s(X_s))=P(\theta_s(X_t))$, H-score measures the transferability of $\theta_s$ with respect to the target task as
% suppose we have a target data matrix $X_t \in \mathbb{R}^{n \times d}$ (n is the total number of data samples and d is the dimension of a data sample), corresponding label $Y_t$, and a source model $\theta_s$ pre-trained on a source task $T_s$, we need to evaluate the source model $\theta_s$ on target data $X_t$ to get target features $\theta_s(X_t)$. Then the H-score of $\theta_s$ with respect to a target task is: %with joint probability $P(X_t, Y_t)$ is:
\begin{equation}
\begin{small}
    \mathcal{H}(\theta_s,X_t,Y_t) = \mathrm{tr}(cov(\theta_s(X_t))^{-1}cov(\mathbb{E}_{P_{X_t|Y_t}}[\theta_s(X_t)|Y_t])),
\end{small}
\end{equation} where the covariance ($cov$) and the expectation are estimated empirically from data \cite{bao2019information}. H-score is originally designed to handle classification tasks, therefore we adapt it to segmentation tasks by considering each pixel in the image as an individual classification task and calculating an H-score for each of them. Then the final H-score is the arithmetic mean of all pixel-wise scores:
\begin{equation}
\begin{small}
\textit{H-score}(T_s, T_t) =  \frac{1}{N}\sum_{j=1}^{N}\mathcal{H}(\theta^j_s,X_t, Y_t^{ j}),
\end{small}
\end{equation}
where $N$ is the total number of pixels in the segmentation label image, $\theta^j_s$ represents the source model feature mapping corresponding to the $j$th pixel, and $Y^j_t$ represents the segmentation label of the $j$th pixel. %In p and $\theta_s(X_t)$ is the feature embedding of a pre-trained segmentation model on target data. 
%{\color{red} [to replace]
%:
%\begin{equation}
%\begin{small}
%\textit{H-score}(T_s, T_t) = \frac{1}{n}\sum_{i=1}^{n}\frac{1}{N}\sum_{j=1}^{N}\mathcal{H}(f_t^i, y_t^{i, j}),
%\end{small}
%\end{equation}
%where $N$ is the total number of pixels in the label, $f_t^i$ is the extracted feature of the $ith$ image, and $y_t^{i, j}$ is the label for the $jth$ pixel in the $ith$ image.} %More details of H-score can be found in \cite{bao2019information}. 
We choose H-score because it is highly efficient and effective when the segmentation tasks have similar input distributions but different RoIs.
%To compute OTCE, similarly, the pre-trained source model $\theta_s$ is used to extract features from the source dataset $D_s$ and the target dataset $D_t$. \textcolor{red}{To handle segmentatition tasks}, the feature vector $f$ and the label $y$ of each pixel are used to construct pixel-wise datasets, denoted as $D^{pix}_s=\{(f^i_s,y^i_s)\}^{N_s}_{i=1}$ for the source task and $D^{pix}_t=\{(f^i_t,y^i_t)\}^{N_t}_{i=1}$ for the target task, where $N_s=m \times H \times W$ and $N_t=n \times H \times W$, representing the number of pixels of all samples in the source and target datasets, respectively. Then the optimal transport (OT) problem can be defined as: 

The OTCE score is a versatile transferability metric for both the cross-domain and the cross-task transfer scenarios. To compute OTCE, we first use the pre-trained source model $\theta_s$ to produce the pixel-wise feature sets $D_s=\{(v^i_s,y^i_s)\}^{N_s}_{i=1} $ and $D_t=\{(v^i_t,y^i_t)\}^{N_t}_{i=1}$ for the source and target tasks respectively, where $v_d^i$, $y_d^i$, and $N_d$ denotes the pixel-wise feature vector, pixel-wise label, and the total number of pixels in all images of task $d\in\{s,t\}$, respectively. %, and $N_s, N_t$ represent the number of pixels for the source and target images respectively. 
Next, we find the optimal coupling matrix of size $N_s\times N_t$ between source and target features by solving the following  regularized Optimal Transport (OT) problem:
\begin{equation}
\begin{small}
\textit{OT}(D_s, D_t)  \triangleq  \mathop{\min}\limits_{\pi \in \Pi(D_s, D_t)}  \sum_{i,j=1}^{N_s,N_t}  ||v^i_s-v^j_t||^2_2\pi_{ij} + \epsilon H(\pi),
\end{small}
\end{equation}
where %$c(\cdot, \cdot) = \| \cdot - \cdot \| ^2_2$ is the cost metric,
%$\pi$ is the coupling matrix of size $N_s\times N_t$, and 
$H(\pi)=-\sum_{i=1}^{N_s} \sum_{j=1}^{N_t} \pi_{ij}\log\pi_{ij}$ is the entropic regularizer with $\epsilon=0.1$. % Solve this OT problem to obtain an optimal coupling matrix $\pi^*$. 
Since the optimal coupling matrix $\pi^*$ represents the empirical joint probability distribution of source and target pixel-wise features, under mild assumptions, the empirical joint probability distribution of source and target labels %as well as 
%the marginal probability distribution of source labels 
can be written as
\begin{equation}
\begin{small}
\hat{P}(y_s,y_t) = \sum_{i,j: y^i_s=y_s, y^j_t=y_t} \pi_{ij}^*.
\end{small}
\end{equation}
%\begin{equation}
%\begin{small}
%\hat{P}(y_s) = \sum_{y_t \in \mathcal{Y}_t} \hat{P}(y_s,y_t),
%\end{small}
%\end{equation}
Finally, the OTCE score can be computed as the negative conditional entropy between the source and target labels:
\begin{equation}
\begin{small}
\begin{aligned}
\textit{OTCE}(T_s, T_t) & = - H(Y_t|Y_s) \\
& = \sum_{y_t \in \mathcal{Y}_t} \sum_{y_s \in \mathcal{Y}_s} \hat{P}(y_s,y_t)\log \frac{\hat{P}(y_s,y_t)}{  \sum_{y_t \in \mathcal{Y}_t}\hat{P}(y_s,y_t)  },
\end{aligned}
\end{small}
\end{equation}
where $\mathcal{Y}_s$ and $\mathcal{Y}_t$ denote the source and target pixel-wise label space. More details of OTCE for semantic segmentation can be found in \cite{tan2021transferability}.

As an ablation study of our source selection framework, a baseline  approach   (Path 2 in Fig. \ref{fig:sources_selection_framwork}) is directly computing the transferability estimation metric on all source tasks without considering the image characteristics of tasks.

\section{Experimental Settings and Results}
\subsection{Datasets}
We perform experiments on three publicly available brain MRI segmentation datasets: FeTS 2021 \cite{pati2021federated,reina2021openfl,bakas2017advancing} for brain tumor segmentation, iSeg-2019 \cite{sun2021multi} for brain matter segmentation, and WMH \cite{kuijf2019standardized} for white matter hyperintensities segmentation.

For each sample in FeTS 2021 dataset, volumes of 4 modalities are available, including T1-weighted (T1), T2-weighted (T2), Fluid-Attenuated Inversion Recovery (FLAIR), and T1-Weighted Contrast-Enhanced (T1CE). The volume size is $240 \times 240 \times 155$. Corresponding labels of edematous tissue (ED), enhancing tumor (ET), and necrotic tumor core (NCR) are manually segmented by clinical experts. For each sample in iSeg-2019 dataset, volumes of 2 modalities are available, including T1 and T2. The volume size is $144 \times 192 \times 256$. Corresponding labels of white matter (WM), gray matter (GM), and cerebrospinal fluid (CSF) are manually segmented by clinicians. The WMH dataset consists of 60 brain MRI volumes of FLAIR modality with manual annotations of white matter hyperintensities from three different institutes, namely, VU Amsterdam (A), NUHS Singapore (S), and UMC Utrecht (U). Volume sizes are $132\times256\times83$, $256\times232\times48$, and $240\times240\times48$ for the three institutes, respectively. Corresponding labels of white matter hyperintensities (WMH) are manually segmented by clinical experts.

\begin{figure}[t]
  \centering
  \includegraphics[width=0.6\linewidth]{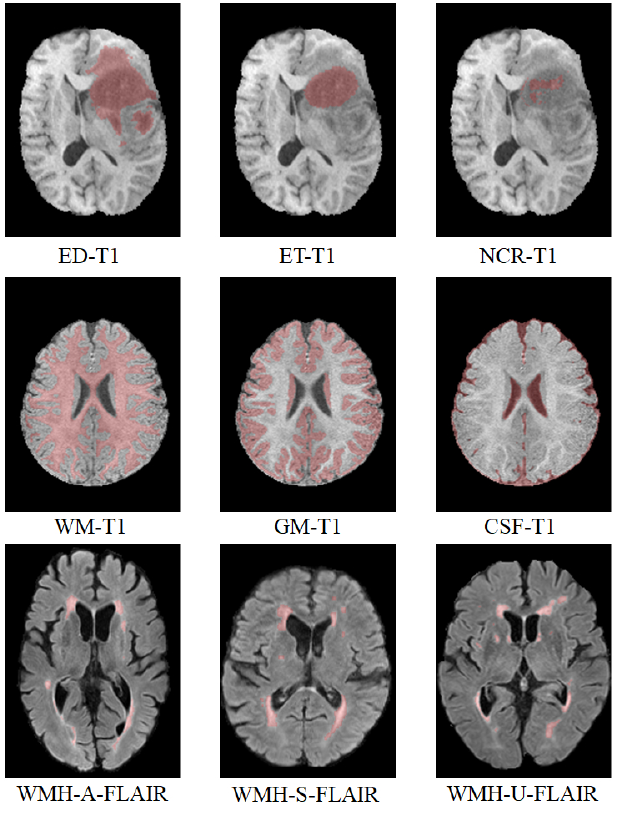}
  \vspace{-0.4cm}
  \caption{Examples of images from FeTS 2021, iSeg-2019, and WMH datasets with their corresponding labels. In FeTS 2021 dataset, e.g., ``ED-T1" denotes edematous tissue segmentation on T1 modality. In iSeg-2019 dataset, e.g., ``WM-T1" denotes white matter segmentation on T1 modality. In WMH dataset, e.g., ``WMH-A-FLAIR" denotes white matter hyperintensities segmentation on FLAIR modality from institute A.}
  \label{fig:datasets}
  \vspace{-0.4cm}
\end{figure}

To provide enough tasks for experimental analysis, we reorganize these three datasets into a collection of binary segmentation tasks on every available modality. Examples of images from these three datasets with their corresponding labels are visualized in Fig. \ref{fig:datasets}.

\subsection{Model Architecture and Transfer Learning Strategy}
Since the goal of this work is to investigate the source selection problem rather than trying to achieve state-of-the-art performance on medical image segmentation tasks, we use the same model architecture for all experiments presented in this paper, a classic 2D U-Net \cite{ronneberger2015u}. As for the transfer learning strategy, we follow the most common way which is pre-training the model on a source task and fine-tuning it on a target task. During the fine-tuning stage, the encoder is frozen and only the parameters of the decoder are updated. See the supplementary materials for details on the model architecture and the training process.

% \vspace{-0.15cm}
\subsection{Verification of Modality Analysis}
In Modality Analysis, we select source tasks having the same modality as the target task as the candidate sources (Subset 1). To verify the correctness of this design, we conduct experiments using source and target tasks from the multi-modal datasets FeTS 2021 and iSeg-2019. 
\begin{table}[t]
\centering
\caption{Comparisons of Cross-Task Transfer and Cross-Modality Transfer on FeTS 2021 (Top) and iSeg-2019 (Bottom) Datasets.}
\label{tab:modality_analysis}
\begin{tabular}{ccc}
\toprule
\textbf{Target}          & \textbf{Source$^{\mathrm{a}}$}               & \textbf{Average Dice$^{\mathrm{b}}$}                          \\ \midrule
\multirow{2}{*}{ET-T1CE}    & {\color[HTML]{009901} ET-T1, ET-T2, ET-FLAIR}     &     0.755                  \\ 
                            & {\color[HTML]{3531FF} ED-T1CE, NCR-T1CE}  &  {\color[HTML]{FF0000} 0.821}\\ \midrule
\multirow{2}{*}{ED-T1CE}    & {\color[HTML]{009901} ED-T1, ED-T2, ED-FLAIR}     &    0.731                   \\
                            & {\color[HTML]{3531FF} ET-T1CE, NCR-T1CE}   & {\color[HTML]{FF0000} 0.786}\\\midrule
\multirow{2}{*}{NCR-T1CE}   & {\color[HTML]{009901} NCR-T1, NCR-T2, NCR-FLAIR}    &    0.726                     \\
                            & {\color[HTML]{3531FF} ET-T1CE, ED-T1CE}   & {\color[HTML]{FF0000} 0.782} \\\midrule\midrule
\multirow{2}{*}{WM-T1}      & {\color[HTML]{009901} WM-T2}    &    0.864                   \\
                            & {\color[HTML]{3531FF} GM-T1, CSF-T1}   & {\color[HTML]{FF0000} 0.877} \\\midrule
\multirow{2}{*}{GM-T1}      & {\color[HTML]{009901} GM-T2}    &    0.881                   \\
                            & {\color[HTML]{3531FF} WM-T1, CSF-T1}   & {\color[HTML]{FF0000} 0.892} \\\midrule
\multirow{2}{*}{CSF-T1}      & {\color[HTML]{009901} CSF-T2}    &    {\color[HTML]{FF0000} 0.935}                   \\
                            & {\color[HTML]{3531FF} WM-T1, GM-T1}   &  0.934\\
\bottomrule
\multicolumn{3}{l}{$^{\mathrm{a}}$Green/Blue: cross-modality/task transfer.}\\
\multicolumn{3}{l}{$^{\mathrm{b}}$Red: better transfer performance.}
\end{tabular}
\vspace{-0.5cm}
\end{table}

As shown in Table \ref{tab:modality_analysis}, transferring from different tasks under the same modality (sources in blue) outperforms transferring from the same task under different modality (sources in green) in almost all trials (5 out of 6, the scores of the only exception are very close) with a significant margin (average Dice scores in red). Such a finding suggests that for the common transfer learning strategy of pre-training and fine-tuning, matching different source and target data modalities is harder than re-learning a new task within the same modality.

\subsection{Verification of RoI Analysis}
In RoI Analysis, within Subset 1, we select source tasks whose RoI shapes are more similar to that of the target task by calculating the shape similarity using SSIM, and then generate Subset 2. To verify this choice, we conduct experiments using source and target tasks from all three datasets. Specifically, our experiment includes both the setting when the source and target tasks are from the same dataset (same-dataset), and when they are from different datasets (cross-dataset) to show the effectiveness and generalizability of our framework.

\begin{table*}[t]
	\centering
\begin{minipage}{\linewidth}
\centering
\setlength\tabcolsep{4pt} 
\footnotesize
\caption{Analysis of Relationship between RoI Shape Similarity and Transfer Performance on FeTS 2021 (Top) and iSeg-2019 (Bottom) Datasets.}
\label{tab:roi_same}
\begin{tabular}{cccc |cccc |cccc}
\toprule
\textbf{Target}             & \textbf{Source} & \multicolumn{1}{l}{\textbf{Dice$^{\mathrm{a}}$}} & \multicolumn{1}{l}{\textbf{RoI-Sim$^{\mathrm{b}}$}} &
\textbf{Target}             & \textbf{Source} & \multicolumn{1}{l}{\textbf{Dice$^{\mathrm{a}}$}} & \multicolumn{1}{l}{\textbf{RoI-Sim$^{\mathrm{b}}$}} &
\textbf{Target}             & \textbf{Source} & \multicolumn{1}{l}{\textbf{Dice$^{\mathrm{a}}$}} & \multicolumn{1}{l}{\textbf{RoI-Sim$^{\mathrm{b}}$}} \\ \midrule
\multirow{2}{*}{ED-T1} & ET-T1 &  {\color[HTML]{FF0000} 0.757} & {\color[HTML]{FF0000} 0.987} &
\multirow{2}{*}{ED-T2} & ET-T2 &  {\color[HTML]{FF0000} 0.811} & {\color[HTML]{FF0000} 0.987} &
\multirow{2}{*}{ED-FLAIR} & ET-FLAIR &  {\color[HTML]{FF0000} 0.773} & {\color[HTML]{FF0000} 0.987}\\
  & NCR-T1 & 0.738 &  0.984 &  & NCR-T2 & 0.802 & 0.984 &  & NCR-FLAIR & 0.760 & 0.984\\\midrule
\multirow{2}{*}{ET-T1} & ED-T1 &  {\color[HTML]{FF0000} 0.693} & {\color[HTML]{FF0000} 0.987} &
\multirow{2}{*}{ET-T2} & ED-T2 &  {\color[HTML]{FF0000} 0.703} & {\color[HTML]{FF0000} 0.987} &
\multirow{2}{*}{ET-FLAIR} & ED-FLAIR &  {\color[HTML]{FF0000} 0.681} & {\color[HTML]{FF0000} 0.987}\\
  & NCR-T1 & 0.672 &  0.985 &  & NCR-T2 & 0.660 & 0.985 &  & NCR-FLAIR & 0.613 & 0.985 \\\midrule
\multirow{2}{*}{NCR-T1} & ED-T1 & 0.585 & 0.984 &
\multirow{2}{*}{NCR-T2} & ED-T2 & {\color[HTML]{FF0000} 0.598} & 0.984 &
\multirow{2}{*}{NCR-FLAIR} & ED-FLAIR & {\color[HTML]{FF0000} 0.558} & 0.984\\
  & ET-T1 & {\color[HTML]{FF0000} 0.602} & {\color[HTML]{FF0000} 0.985} &  & ET-T2 & 0.578 & {\color[HTML]{FF0000} 0.985} &
  & ET-FLAIR & 0.550 & {\color[HTML]{FF0000} 0.985}
  \\
  \midrule
   \midrule
\multirow{2}{*}{WM-T1} & GM-T1 &  {\color[HTML]{FF0000} 0.885} & {\color[HTML]{FF0000} 0.844} &
\multirow{2}{*}{WM-T2} & GM-T2 &  {\color[HTML]{FF0000} 0.852} & {\color[HTML]{FF0000} 0.844} &
\multirow{2}{*}{-} & - &  - & -\\
  & CSF-T1 & 0.868 &  0.824 &  & CSF-T2 & 0.815 & 0.824 &  & - & - & -\\\midrule
\multirow{2}{*}{GM-T1} & WM-T1 &  {\color[HTML]{FF0000} 0.893} & {\color[HTML]{FF0000} 0.844} &
\multirow{2}{*}{GM-T2} & WM-T2 &  {\color[HTML]{FF0000} 0.879} & {\color[HTML]{FF0000} 0.844} &
\multirow{2}{*}{-} & - &  - & -\\
  & CSF-T1 & 0.891 &  0.838 &  & CSF-T2 & 0.863 & 0.838 &  & - & - & - \\\midrule
\multirow{2}{*}{CSF-T1} & WM-T1 & 0.925 & 0.824 &
\multirow{2}{*}{CSF-T2} & WM-T2 & 0.909 & 0.824 &
\multirow{2}{*}{-} & - & - & -\\
  & GM-T1 & {\color[HTML]{FF0000} 0.942} & {\color[HTML]{FF0000} 0.838} &  & GM-T2 & {\color[HTML]{FF0000} 0.916} & {\color[HTML]{FF0000} 0.838} &
  & - & - & -\\
\bottomrule
\multicolumn{8}{l}{$^{\mathrm{a}}$Red: better transfer performance. $^{\mathrm{b}}$Red: stronger RoI shape similarity.}
\end{tabular}
\vspace{-0.4cm}
\end{minipage}
\end{table*}

\begin{table}[t]
\centering
\caption{Analysis of Relationship between RoI Shape Similarity and Transfer Performance across FeTS 2021 and WMH Datasets.}
\label{tab:roi_cross}
\begin{tabular}{cccc}
\toprule
\textbf{Target}          & \textbf{Source}             & \textbf{Dice$^{\mathrm{a}}$}                & \textbf{RoI-Sim$^{\mathrm{b}}$}                \\ \midrule
                              & {\color[HTML]{333333} NCR-FLAIR} & {\color[HTML]{FF0000} 0.650} & {\color[HTML]{FF0000} 0.979} \\
                              & {\color[HTML]{333333} ET-FLAIR}  & 0.644                        & 0.978                        \\
\multirow{-3}{*}{WMH-U-FLAIR} & {\color[HTML]{333333} ED-FLAIR}  & 0.578                        & 0.967                        \\ \midrule
                              & {\color[HTML]{333333} NCR-FLAIR} & {\color[HTML]{FF0000} 0.542} & {\color[HTML]{FF0000} 0.982} \\
                              & {\color[HTML]{333333} ET-FLAIR}  & {\color[HTML]{FF0000} 0.542} & 0.981                        \\
\multirow{-3}{*}{WMH-S-FLAIR} & {\color[HTML]{333333} ED-FLAIR}  & 0.531                        & 0.970                        \\ \midrule
                              & {\color[HTML]{333333} NCR-FLAIR} & {\color[HTML]{FF0000} 0.572} & {\color[HTML]{FF0000} 0.985} \\
                              & {\color[HTML]{333333} ET-FLAIR}  & 0.538                        & 0.984                        \\
\multirow{-3}{*}{WMH-A-FLAIR} & {\color[HTML]{333333} ED-FLAIR}  & 0.547                        & 0.973                        \\ \bottomrule
\multicolumn{4}{l}{$^{\mathrm{a}}$Red: better transfer performance.}\\
\multicolumn{4}{l}{$^{\mathrm{b}}$Red: stronger RoI shape similarity.}
\end{tabular}
\vspace{-0.4cm}
\end{table}

As shown in Table \ref{tab:roi_same} and \ref{tab:roi_cross}, we first compute the RoI shape similarity score (RoI-Sim) between the source and the target task of the same modality. Then, we also perform transfer learning experiments to obtain the ground truth transfer accuracy between them. According to the results, we can conclude that in both same-dataset (Table \ref{tab:roi_same}) and cross-dataset (Table \ref{tab:roi_cross}) settings, when the RoI shape similarity between the source and the target task is stronger (the RoI-Sim score in red), the transfer performance is often (16 out of 18 trials) better (the Dice score in red). Such a finding suggests that within the same modality, SSIM can serve as an indicator to rank the performances of transfer learning from different source tasks to the target task.

% \vspace{-0.2cm}
\subsection{Results of Source Selection}
%Here we present experiements that evaluates the proposed source selection framework.
For the source selection experiments, we use FeTS 2021 dataset. This dataset is further split into 22 partitions by the provider, according to different institutions and information extracted from images. Thus, each partition can be seen as an individual domain. Here, we additionally denote a  task by ``Task-Partition-Modality", e.g., ``ET-14-T1" represents the task of enhancing tumor segmentation on T1 modality using data from partition 14. In total, 16 source tasks (ED/NCR-13/14/17/18-T1/T2) and 2 target tasks (ET-22-T2 and ET-20-T1) are used to conduct two groups of source selection experiments. The ground truth transfer learning results and the calculated transferability results on two target tasks  are  shown  in Table \ref{tab:et_22_t2} and Table \ref{tab:et_20_t1}.
\begin{table}[t]
\centering
\caption{Transfer Learning and Transferability Estimation Results on the Target Task of ET-22-T2.}
\label{tab:et_22_t2}
\begin{tabular}{ccccc}
\toprule
\textbf{Target}        & \textbf{Source$^{\mathrm{a}}$}                  & \textbf{Dice} & \textbf{H-score} & \textbf{OTCE} \\ \midrule
                            & ED-14-T1                              & 0.664         & -0.0380          & -0.0395        \\
                            & {\color[HTML]{FF0000} {\underline {ED-14-T2}}} & 0.703         & 0.1887           & -0.0226        \\
                            & NCR-14-T1                             & 0.646         & 0.8990           & -0.0395        \\
                            & {\underline {NCR-14-T2}}                       & 0.660         & 0.5140           & -0.0383        \\
                            & ED-13-T1                              & 0.657         & 0.4142           & -0.0407        \\
                            & {\color[HTML]{FF0000} {\underline {ED-13-T2}}} & 0.695         & 1.4031           & -0.0356        \\
                            & NCR-13-T1                             & 0.628         & 5.0050           & -0.0407        \\
                            & {\underline {NCR-13-T2}}                       & 0.683         & 10.5247          & -0.0401        \\
                            & ED-17-T1                              & 0.697         & 0.3525           & -0.0435        \\
                            & {\color[HTML]{FF0000} {\underline {ED-17-T2}}} & 0.708         & 1.3327           & -0.0389        \\
                            & NCR-17-T1                             & 0.612         & 1.5211           & -0.0436        \\
                            & {\underline {NCR-17-T2}}                       & 0.681         & 6.6535           & -0.0433        \\
                            & ED-18-T1                              & 0.675         & 0.1070           & -0.0435        \\
                            & {\color[HTML]{FF0000} {\underline {ED-18-T2}}} & 0.707         & 0.2776           & -0.0273        \\
                            & NCR-18-T1                             & 0.664         & 0.9038           & -0.0436        \\
\multirow{-16}{*}{ET-22-T2} & {\underline {NCR-18-T2}}                       & 0.666         & 2.2038           & -0.0394        \\
\bottomrule
\multicolumn{5}{l}{$^{\mathrm{a}}$Underlined sources: Subset 1. Sources in red: Subset 2.}
\end{tabular}
\vspace{-0.5cm}
\end{table}

\begin{table}[t]
\centering
\caption{Transfer Learning and Transferability Estimation Results on the Target Task of ET-20-T1.}
\label{tab:et_20_t1}
\begin{tabular}{ccccc}
\toprule
\textbf{Target}        & \textbf{Source$^{\mathrm{a}}$}                   & \textbf{Dice} & \textbf{H-score} & \textbf{OTCE} \\ \midrule
                            & {\color[HTML]{FF0000} {\underline {ED-14-T1}}}  & 0.636         & 1.5433           & -0.0320        \\
                            & {\color[HTML]{333333} ED-14-T2}        & 0.609         & 0.2268           & -0.0330        \\
                            & {\color[HTML]{333333} {\underline {NCR-14-T1}}} & 0.560         & 3.2383           & -0.0325        \\
                            & {\color[HTML]{333333} NCR-14-T2}       & 0.627         & 2.5139           & -0.0330        \\
                            & {\color[HTML]{FF0000} {\underline {ED-13-T1}}}  & 0.593         & 1.7564           & -0.0333        \\
                            & {\color[HTML]{333333} ED-13-T2}        & 0.636         & 1.3293           & -0.0347        \\
                            & {\color[HTML]{333333} {\underline {NCR-13-T1}}} & 0.498         & 2.4574           & -0.0342        \\
                            & {\color[HTML]{333333} NCR-13-T2}       & 0.610         & 6.5037           & -0.0346        \\
                            & {\color[HTML]{FF0000} {\underline {ED-17-T1}}}  & 0.680         & 2.5901           & -0.0351        \\
                            & {\color[HTML]{333333} ED-17-T2}        & 0.571         & -3.2459          & -0.0363        \\
                            & {\color[HTML]{333333} {\underline {NCR-17-T1}}} & 0.581         & 25.7285          & -0.0361        \\
                            & {\color[HTML]{333333} NCR-17-T2}       & 0.532         & 40.6843          & -0.0363        \\
                            & {\color[HTML]{FF0000} {\underline {ED-18-T1}}}  & 0.613         & 0.0901           & -0.0357        \\
                            & {\color[HTML]{333333} ED-18-T2}        & 0.616         & 0.3164           & -0.0355        \\
                            & {\color[HTML]{333333} {\underline {NCR-18-T1}}} & 0.632         & 0.2743           & -0.0361        \\
\multirow{-16}{*}{ET-20-T1} & {\color[HTML]{333333} NCR-18-T2}       & 0.637         & 1.6508           & -0.0362        \\ \bottomrule
\multicolumn{5}{l}{$^{\mathrm{a}}$Underlined sources: Subset 1. Sources in red: Subset 2.}
\end{tabular}
\vspace{-0.5cm}
\end{table}

The result of source selection is a ranking of source tasks according to their transfer performance on a given target task. The \textbf{ground truth ranking} is obtained by sorting the Dice scores after fine-tuning each source task on a given target task. A higher Dice score indicates better transferability. The \textbf{baseline ranking prediction} is obtained by directly computing and sorting H-scores or OTCE scores on all source tasks (Path 2 in Fig. \ref{fig:sources_selection_framwork}). A higher H-score or OTCE score indicates better transferability. The \textbf{ranking prediction proposed by our framework} is obtained through combining prior knowledge with transferability estimation metrics, as indicated by Path 1 in Fig. \ref{fig:sources_selection_framwork}. We take Table \ref{tab:et_22_t2} as an example to illustrate how to obtain the ranking prediction with our proposed framework. Given a target task of ET-22-T2, we notice that this task is under the T2 modality. According to our modality analysis in Section \ref{sec:modality}, we should select those 8 source tasks under the same modality as the target task (underlined in Table \ref{tab:et_22_t2}). This procedure forms   Subset 1. Next, according to our RoI analysis in Section \ref{sec:roi}, within the T2 modality, the RoI shape similarity between ED and ET estimated by SSIM is higher than that between ED and NCR, thus we should select those 4 source tasks (colored in red in Table \ref{tab:et_22_t2}) of ED segmentation rather than NCR segmentation. This procedure forms Subset 2. Finally, we apply the analytical transferability estimation metric on source tasks in Subset 2 and obtain their predicted ranking. As for Table \ref{tab:et_20_t1}, the source selection procedure using our proposed framework is similar.

The performance of source selection methods is often evaluated by comparing the difference between the ground truth transfer performance ranking and the predicted transfer performance ranking. Here, we use Spearman’s footrule \cite{diaconis1977spearman} to quantify the difference between the two rankings. More details on Spearman’s footrule can be found in the supplementary materials.
\begin{table}[t]
\centering
\caption{Evaluation of Source Selection Performance with/without Prior Knowledge.}
\label{tab:footrule}
\begin{tabular}{cccccc}
\toprule
\textbf{Target}       & \textbf{Method$^{\mathrm{a}}$} & \textbf{Top 1$^{\mathrm{b}}$}           & \textbf{Top 2$^{\mathrm{b}}$}            & \textbf{Top 3$^{\mathrm{b}}$}            & \textbf{Top 4$^{\mathrm{b}}$}            \\ \midrule
                           & H-score w/o PK   & 5                        & 10                        & 22                        & 27                        \\
                           & H-score w/ PK    & {\color[HTML]{FF0000} 4} & {\color[HTML]{FF0000} 5}  & {\color[HTML]{FF0000} 6}  & {\color[HTML]{FF0000} 7}  \\ \cmidrule{2-6} 
                           & OTCE w/o PK      & 2                        & 2                         & 4                         & 12                        \\
\multirow{-4}{*}{ET-22-T2} & OTCE w/ PK       & 2                        & 2                         & 4                         & {\color[HTML]{FF0000} 7}  \\ \midrule
                           & H-score w/o PK   & 14                       & 24                        & 30                        & 40                        \\
                           & H-score w/ PK    & {\color[HTML]{FF0000} 0} & {\color[HTML]{FF0000} 9}  & {\color[HTML]{FF0000} 9}  & {\color[HTML]{FF0000} 13} \\ \cmidrule{2-6}  
                           & OTCE w/o PK      & 2                        & 14                        & 17                        & 23                        \\
\multirow{-4}{*}{ET-20-T1} & OTCE w/ PK       & 2                        & {\color[HTML]{FF0000} 11} & {\color[HTML]{FF0000} 13} & {\color[HTML]{FF0000} 17} \\
\bottomrule
\multicolumn{6}{l}{$^{\mathrm{a}}$PK: prior knowledge.}\\
\multicolumn{6}{l}{$^{\mathrm{b}}$Top 1-4: required number of selected sources. Red: better performance.}
\end{tabular}
\vspace{-0.5cm}
\end{table}
The performance evaluation on selecting the top 1-4 source tasks is shown in Table \ref{tab:footrule}. %It can be observed that 
For both target tasks and under all top 1-4 source selection settings, when following our proposed prior knowledge guided and transferability based framework, the difference between the predicted ranking and the ground truth ranking is reduced. This suggests that prior knowledge about the medical image segmentation tasks, including modality and RoI characteristics, can indeed improve the current state-of-the-art transferability metrics' ability to successfully select source tasks with better transfer performance. Besides, these results also reveal that current transferability estimation metrics are not sufficient to handle the large gaps between source and target tasks and thus require further refinement, particularly in the medical domain.

\section{Conclusion}
We propose a prior knowledge guided and transferability based framework to tackle the source selection problem in transfer learning for brain image segmentation tasks. We are the first to apply state-of-the-art   transferability estimation metrics to the medical image segmentation domain. Different from the common procedure that directly applies these metrics, our framework further considers the prior knowledge of the given source and target tasks when selecting sources. Specifically, we perform modality analysis and RoI analysis to select a subset of source tasks and then only compute the metric within this subset. Modality analysis shows that transferring from different tasks under the same modality is better than transferring from the same task under different modalities. RoI Analysis shows that stronger RoI shape similarity between the source and the target task often leads to better transfer performance. Consequently, by incorporating image characteristics of modality difference and RoI similarity into the framework, source selection experiments suggest that the performance of transferability estimation metrics like H-score and OTCE on the source selection problem can be significantly enhanced. We envision that this framework can be used on other medical image datasets besides brain images.

\bibliographystyle{IEEEbib}
\bibliography{refs}

% \end{document}
% \documentclass[conference]{IEEEtran}
% \IEEEoverridecommandlockouts
% The preceding line is only needed to identify funding in the first footnote. If that is unneeded, please comment it out.
% \usepackage{booktabs}
% \usepackage{multirow}
% \usepackage{flushend}
% \usepackage{cite}
% \usepackage{amsmath,amssymb,amsfonts}
% \usepackage{algorithmic}
% \usepackage{graphicx}
% \usepackage{textcomp}
% \usepackage{xcolor}
% \def\BibTeX{{\rm B\kern-.05em{\sc i\kern-.025em b}\kern-.08em
%     T\kern-.1667em\lower.7ex\hbox{E}\kern-.125emX}}
% \begin{document}

% \makeatletter

% \def\ps@IEEEtitlepagestyle{%
%   \def\@oddfoot{\mycopyrightnotice}%
%   \def\@evenfoot{}%
% }
% \def\mycopyrightnotice{%
%   {\footnotesize 978-1-6654-6819-0/22/\$31.00 \copyright2022 IEEE\hfill}
%   \gdef\mycopyrightnotice{}
% }

% \title{Finding the Most Transferable Tasks for Brain Image Segmentation - Supplementary Materials\thanks{This research is funded by Natural Science Foundation of China 62001266.}
% }

% \author{\IEEEauthorblockN{Yicong Li, Yang Tan, Jingyun Yang, Yang Li, Xiao-Ping Zhang}
% \IEEEauthorblockA{\textit{Tsinghua-Berkeley Shenzhen Institute}\\
% \textit{Tsinghua University}\\
% Shenzhen, China\\
% yangli@sz.tsinghua.edu.cn}}

% \maketitle
 
\section{Supplementary Materials}
\subsection{Structural Similarity Index Measure (SSIM)}
SSIM is often used to evaluate the visual similarity between two images. The idea is that natural images often contain highly structural information, i.e., neighboring pixels in natural images have a strong correlation. And such correlation encompasses the structural information of the object in a given environment. The human visual system is very used to extract such structural information from natural images. Therefore, the measurement of similarity given by SSIM is more in line with the perception of human eyes \cite{wang2009mean,sheikh2006statistical} compared to other metrics like peak signal-to-noise ratio (PSNR). Given two images $x$ and $y$, SSIM can be calculated as:
\begin{equation}
  \textit{SSIM}(x,y)=\frac{\left(2\mu_x\mu_y+C_1\right)\left(2\sigma_{xy}+C_2\right)}{\left(\mu_x^2+\mu_y^2+C_1\right)\left(\sigma_x^2+\sigma_y^2+C_2\right)},
\end{equation}
where $\mu_x$ is the average of $x$, $\mu_y$ is the average of $y$, $\sigma_x^2$ is the variance of $x$, $\sigma_y^2$ is the variance of $y$, and $\sigma_{xy}$ is the covariance of $x$ and $y$. $C_1$ and $C_2$ are constants for maintaining stability. Higher SSIM indicates stronger similarity between $x$ and $y$. It ranges from 0 to 1 and when the two images are identical, the value equals 1.

\subsection{Model and Training Details}
\subsubsection{Model Architecture}
\begin{figure}[t]
   \centering
   \includegraphics[width=\linewidth]{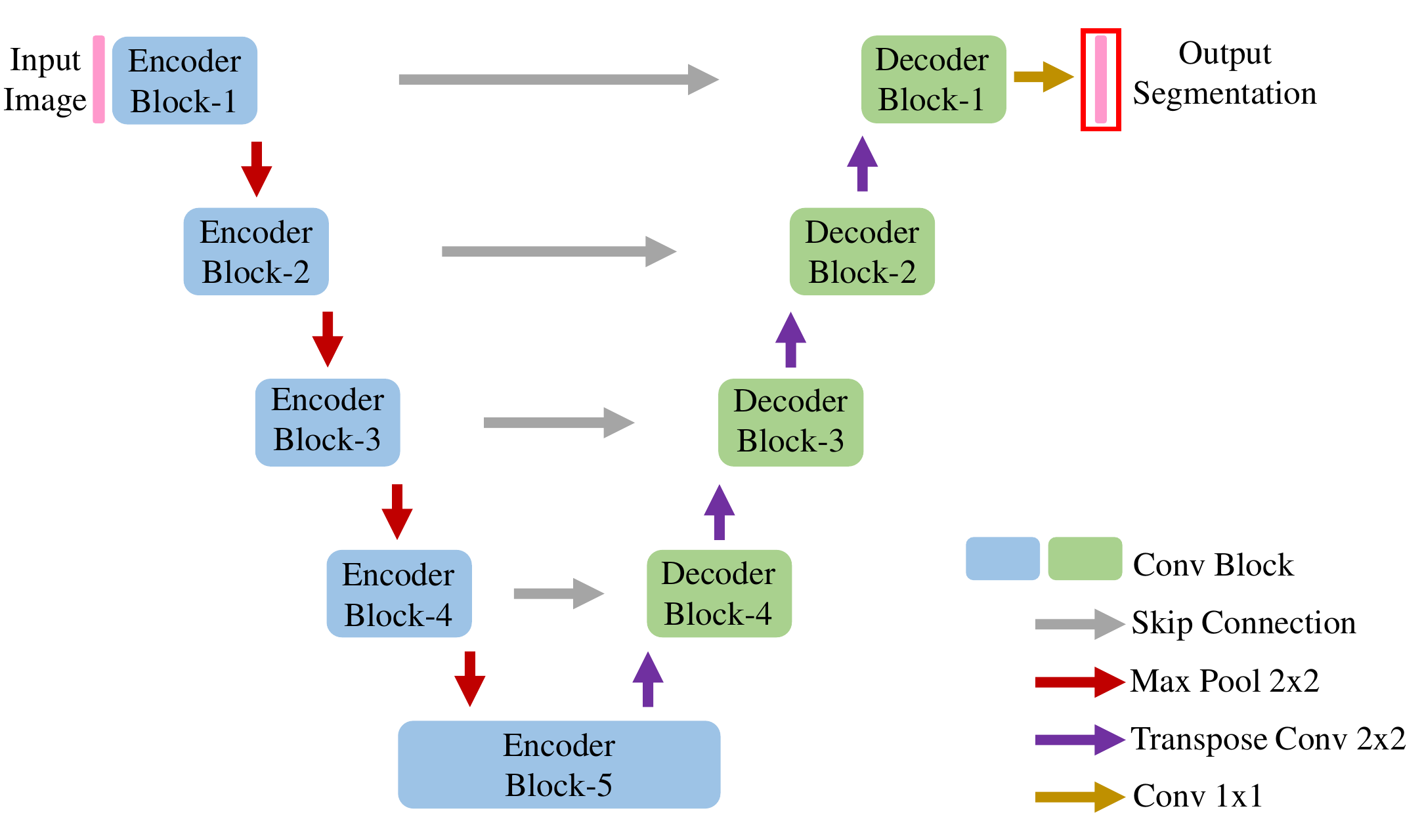}
  \caption{U-Net architecture. Parameters of convolutional blocks in blue are frozen during the fine-tuning stage. Parameters of convolutional blocks in green are updated during the fine-tuning stage. Red box indicates the feature we use to compute the transferability scores. This figure is partially reproduced from \cite{ronneberger2015u}.}
   \label{fig:unet}
  % \vspace{-0.5cm}
 \end{figure}

We use the same model architecture for all experiments presented in this paper, a classic 2D U-Net \cite{ronneberger2015u}, as shown in Fig. \ref{fig:unet}. The model includes an encoder and a decoder. The encoder consists of 5 blocks, each of which contains 2 sub-blocks of $3 \times 3$ convolution, batch normalization, and ReLU activation, followed by a $2 \times 2$ max pooling layer. In the decoder, similar blocks are used, each of which is followed by a $2 \times 2$ transpose convolution layer. The final $1 \times 1$ convolution layer outputs a segmentation map (logits) with 2 channels. During transfer learning, we freeze the encoder and fine-tune the decoder. The final feature map (output segmentation) produced by the decoder is used to compute the transferability score.
 
\subsubsection{Training Configurations}
During the training stage, we use Adam \cite{kingma2014adam} as the optimizer with a learning rate of 1e-4, a weight decay of 5e-5, and a batch size of 20. We set 10,000 iterations for the pre-training stage and 1,000 iterations for the fine-tuning stage. Cross-entropy is used as the loss function and Dice score is chosen as the metric to evaluate the segmentation performance. The model is implemented using Python 3.6.8 and the deep learning framework PyTorch 1.8.0 \cite{paszke2019pytorch}. All experiments are conducted on a CentOS 7.6.1810 system with one GeForce RTX 3090 GPU.

\subsection{Spearman’s Footrule}
Spearman's footrule \cite{diaconis1977spearman} measures the absolute distance between two rankings by calculating how many steps we need to move the elements in the predicted ranking, in order to make it the same as the ground truth ranking.
Formally, given two rankings $A$ and $B$ with the same number (denoted as $N$) of elements, Spearman's footrule is calculated as:
\begin{equation}
\textit{Spearman}(A, B) = \sum_{n=1}^N \lvert A[n] - B[n] \rvert,
\label{eq:spearman}
\end{equation}
For example, if $A = [1, 2, 3]$ and $B = [2, 1, 3]$, then $\textit{Spearman}(A, B) = \lvert 1-2 \rvert + \lvert 2-1 \rvert + \lvert 3-3 \rvert = 2$.

% \bibliographystyle{IEEEbib}
% \bibliography{refs}
% \end{document}

\end{document}